\documentstyle[aps,epsf,twocolumn]{revtex}
\def\Frac#1#2{\frac{\displaystyle{#1}}{\displaystyle{#2}}}
\def\lsim{\raise0.3ex\hbox{$\;<$\kern-0.75em\raise-1.1ex\hbox{$\sim\;$}}}
\def\gsim{\raise0.3ex\hbox{$\;>$\kern-0.75em\raise-1.1ex\hbox{$\sim\;$}}}
\begin{document}
\draft
%\title{Complete of a low CP asymmetry in extended quark models}
\title{Tree--level FCNC in the B system: from CP asymmetries to rare decays}
\author{G. Barenboim}
\address{Theory Division, CERN, CH-1211 Geneva 23, Switzerland}
\author{F.J. Botella and O. Vives}
\address{Departament de F\'{\i}sica Te\`orica and IFIC, \\
Universitat de Val\`encia-CSIC, E-46100, Burjassot, Spain}
\date{\today}
\preprint{CERN-TH/2000-376, FTUV-00-12-15, IFIC-00-82}
\maketitle
\begin{abstract}
Tree--level Flavor--Changing Neutral Currents (FCNC) are characteristic
of models with extra vector--like quarks. These new
couplings can strongly modify the $B^0$ CP asymmetries without conflicting
with low--energy constraints. In the light of a low CP asymmetry 
in $B\to J/\psi\ K_{S}$, we discuss the implications of these 
contributions. We find that even these low 
values can be easily accommodated in these models. 
Furthermore, we show that the new data from 
$B$ factories tend to favor an ${\cal{O}}(20)$ enhancement of the 
$b\rightarrow d l \bar{l}$ transition over the SM expectation.
\end{abstract}
\pacs{12.60.-i, 12.15.Ff, 11.30.Er, 13.25.Hw, 13.25.Es}

The achievements of the Standard Model (SM) \`a la
Cabibbo--Kobayashi--Maskawa (CKM) are really impressive, even in the
flavor and CP violation sectors. It is worth remembering that, within
the Standard Model, it is possible to ``detect'' CP violation 
using purely CP--conserving observables \cite{quico,branco}. This has
been achieved through the combination of $R_{u}=|V_{ub}^{*}V_{ud}|/
|V_{cb}^{*}V_{cd}| $, $|V_{cb}^{*}V_{cd}| $ and $\Delta
m_{B_{d}}$. Furthermore, this CP violation is compatible with
$\varepsilon_{K}$, the measurement of the indirect CP violation in the
kaon system. In fact, taking into account the hadronic uncertainties,
it is hard (today) to say that there is real trouble in the kaon
sector of the SM, even after the inclusion of
$\varepsilon^{\prime}/\varepsilon$ and rare kaon decays. The situation
is slowly changing with the new data in the $B$ sector, after the
Babar and Belle collaborations have started to give results on the
$B\rightarrow J/\psi\ K_{S}$ asymmetry $a_{J/\psi}$. The reported
values to date are: $a_{J/\psi} = 0.34 \pm 0.20 \pm 0.05$
(Babar \cite{babar}), $a_{J/\psi} = 0.58^{+0.32+0.09}_{-0.34-0.19}$
(Belle \cite{belle}) and $a_{J/\psi} = 0.79^{+0.41}_{-0.44}$ (CDF
\cite{CDF}); they correspond to an average value of
$a_{J/\psi}=0.51 \pm 0.18$.  On the other hand, the SM prediction is
\begin{eqnarray}
a_{J/\psi}=\sin \left( 2\beta \right), &\ \ \ \ \ & \beta =\arg \left(
-\Frac{V_{cb}^{\ast }V_{cd}}{V_{tb}^{\ast }V_{td}}\right),
\label{ajksm}
\end{eqnarray}
corresponding to $0.59\leq \sin \left( 2\beta \right) \leq
0.82$, which is certainly outside the $1\sigma $ Babar range but not
outside the world average. This potential discrepancy is at the origin
of several papers \cite{smallajk} studying the implications of a small
$a_{J/\psi}$ in the search of new physics.

In this paper, we analyze the implications of this situation for
a realistic model, obtained with the only addition of an isosinglet
down vector--like quark \cite{vlq} to the SM spectrum. This model
naturally arises, for instance, as the low--energy limit of an $E_{6}$
grand unified theory. At a more phenomenological level, models with
isosinglet quarks provide the simplest self--consistent framework to
study deviations of $3\times 3$ unitarity of the CKM matrix as well as
{ \sl flavor--changing neutral currents (\ FCNC ) at tree level}. In
the rest of the paper, we update the strong low--energy constraints on
the tree--level FCNC couplings, we show that a low CP asymmetry in
$B\rightarrow J/\psi\ K_{S}$ can be easily accommodated within the
model, and we point out other observables, correlated with a low
CP asymmetry, which clearly deviate from their SM values.

The model we discuss has been thoroughly described in
Ref.~\cite{vlq}. The presence of an additional down quark 
implies a $4\times 4$ matrix, $V_{i\alpha }$
($i=u,c,t,4$ , $\alpha =d,s,b,b^{\prime }$), diagonalizing the down
quark mass matrix. For our purpose, the relevant information for
the low--energy physics is encoded in this extended mixing matrix. The
charged currents are unchanged except that $V_{CKM}$ is now the
$3\times 4$ upper submatrix of $V$. However, the distinctive feature
of this model is that FCNC enter the neutral current Lagrangian of
the left--handed down quarks:
\begin{eqnarray}
&{\cal L}_Z = \Frac{g}{2 \cos \theta_W} \left[\bar{u}_{L i}\ \gamma^\mu\ 
u_{L i}\ -\  \bar{d}_{L \alpha}\  U_{\alpha \beta}\ \gamma^\mu\ d_{L \beta}\ - 
\nonumber \right.\\
& \left.\ 2\ \sin^2 \theta_W\ J^\mu_{em} \right]\ Z_{\mu}, \nonumber \\
&U_{\alpha \beta }=\sum_{i=u,c,t}V_{\alpha
i}^{\dagger }V_{i\beta }=\delta _{\alpha \beta }-V_{4\alpha }^{*}V_{4\beta },
\label{ZFCNC}
\end{eqnarray}
where $U_{ds}$, $U_{bs}$ or $U_{bd}=-V_{4b}^{*}V_{4d} \neq 0$ would
signal new physics and the presence of FCNC at tree level. In order to
fully include all the correlations in the analysis below, we use the
following parametrization \cite{quico} for the mixing matrix $V$:
\begin{equation}
V=R_{34}\left( \theta _{34},0\right) R_{24}\left( \theta _{24},\phi
_{3}\right) R_{14}\left( \theta _{14},\phi _{2}\right) V_{CKM}^{SM},
\label{matrixV}
\end{equation}
where $V_{CKM}^{SM}\left( \theta _{12},\theta _{13},\theta _{23},\phi
_{1}\right) $ is $4\times 4$ block diagonal matrix composed of the standard
CKM \cite{ChauKeung,PDG} and a $1 \times 1$ identity in the $(4,4)$
element, and $R_{ij}\left( \theta _{ij},\phi _{k}\right) $ is a
complex rotation between the $i$ and $j$ ``families''. Note
that, in the limit of small new angles, we follow the usual phase conventions.

Charged--current tree--level decays are not affected by new physics at
leading order; we therefore use the PDG constraints \cite{PDG} for
$\left| V_{ud}\right| $, $\left| V_{us}\right| $, $\left|
V_{cd}\right| $, $\left| V_{cs}\right| $, $\left| V_{cb}\right| $ and
$\left| V_{ub}\right| /\left| V_{cb}\right| $. Another
constraint \cite{LuisJoao,Paco,vlq} comes from the $SU(2)_L$ coupling
of the $Z^0$ to $b\overline{b}$. In the SM, this coupling is 
$(V_{CKM}^\dagger \cdot V_{CKM})_{bb}= 1$, but in this model it is modified to 
$U_{bb}=1-\left|V_{4b}\right| ^{2}$; hence, we have \cite{Paco} $\left|
V_{4b}\right| \leq 0.095$. This bound is indeed very important,
because from unitarity it sets the maximum value for any off--diagonal
element in the fourth row and column of $V$.

The next set of constraints involves FCNC processes where new physics
tree--level diagrams compete with the GIM--suppressed one--loop SM
diagrams.  Let us start with the kaon sector. Here we have $Br\left(
K_{L}\rightarrow \mu \overline{\mu }\right) _{SD}$ and
$\varepsilon^{\prime }/\varepsilon$, that are, as shown in 
Ref.~\cite{BurasSilvestrini}, the relevant constraints to restrict
$U_{ds}$. For $Br\left( K_{L}\rightarrow \mu \overline{\mu }\right)
_{SD}$ we have used the equations and bounds of
Ref.~\cite{BurasSilvestrini}, which agree with the long distance
contribution in \cite{PichDumm},
\begin{eqnarray}
&Br\left( K_{L}\rightarrow \mu \overline{\mu }\right) _{SD}=6.32\times
10^{-3}\ \left[ C_{U2Z}\ \mbox{Re}\left( U_{sd}\right) 
\right.\nonumber \\
&\left. - 6.54\times 10^{-5}\ +
Y_{0}\left( x_{t}\right)\ \mbox{Re}\left( \lambda _{t}^{sd}\right)
\right]^{2}\ \leq\ 2.8\times 10^{-9},
\label{kmumu1}
\end{eqnarray}
where $C_{U2Z}=- (\sqrt{2} G_{F}M_{W}^{2}/\pi ^{2})^{-1}\simeq - 92.7$, 
$\lambda _{i}^{ab}=V_{ia}^{\ast }V_{ib}$ and $Y_{0}$ is the Inami--Lim
function \cite{I-L} defined in\cite{BurasH}.
The calculation of $\varepsilon ^{\prime }/\varepsilon$ is more
unsettled, so we have used the equations of Ref.~\cite{BurasSilvestrini},
but with two different hadronic inputs in the parameter $B_{6}^{(1/2)}$: 
\begin{eqnarray}
&\Frac{\varepsilon^{\prime}}{\varepsilon}=\ \beta_{U}\ C_{U2Z}\ \mbox{Im}\left(
U_{sd}\right)\ +\ \beta_{t}\ \mbox{Im}\left(\lambda _{t}^{sd}\right)\nonumber
\\ 
&\beta_{U}\ =\ [1.2\ -\ R_{s}\ r_{Z}\ B_{8}^{(3/2)}] \nonumber \\ 
&\beta_{t}\ =\ \beta _{U}\cdot C_{0}\ -\ 2.3\ +\ R_{s}\ [1.1\ r_{Z}\
B_{6}^{(1/2)}\ +\ 
\nonumber \\
&(1.0\ +\ 0.12\ r_{Z})\ B_{8}^{(3/2)}].
\label{epsilonprima}
\end{eqnarray}
The first analysis uses $B_{6}^{(1/2)}=1\pm 0.2$ as in
Refs.~\cite{BurasSilvestrini,BurasMart}, and this tends to favor the
presence of new physics in $U_{ds}$. The second one uses
$B_{6}^{(1/2)}=1.3\pm 0.5$ in order to incorporate the predictions of
Refs.~\cite{Bertolini,PichPallante}, where inclusion of the
correction from final--state interactions tends to favor the SM
range. Other parameters are taken as in \cite{BurasSilvestrini}. Once
these two bounds are imposed, the theoretically cleaner bound from
$K^{+}\rightarrow \pi ^{+}\nu \overline{\nu }$ is not
relevant \cite{BurasSilvestrini,Nirnew}. For $\varepsilon _{K}$,
the leading--order expression is \cite{BB}:
\begin{eqnarray}
&\varepsilon_{K} =
\Frac{e^{i\pi /4}G_{F}B_{K}F_{K}^{2}m_{K}}{6\Delta m_{K}}\ \mbox{Im}\left\{
-\left( U_{sd}\right)^{2} + \Frac{\alpha }{4\pi \sin ^{2}\theta_{W}}\right. 
\nonumber \\ 
&\left. \left[ 8\sum\limits_{i=c}^{t}\ Y_{0}\left( x_{i}\right)\
\lambda _{i}^{sd}U_{sd}\  - 
\sum\limits_{i,j=c}^{t}\ S_{0}(x_{i},x_{j})\ \lambda
_{i}^{sd}\lambda _{j}^{sd}\right] \right\}
\label{epsilon}
\end{eqnarray}
where $S_{0}$ is another Inami--Lim function \cite{BurasH}. The QCD
corrections are incorporated as in \cite{BurasSilvestrini}. Contrary
to Ref.~\cite{BurasSilvestrini}, the coefficient $Y_{0}\left(
x\right) $ of the linear term in $U_{ds}$ is characteristic of the present
model, therefore the irrelevance of $\varepsilon _{K}$ to constraint
$U_{ds}$ is not fully guaranteed. On average, once $\varepsilon _{K}$
is irrelevant to constrain $U_{ds}$, the contribution to
$\varepsilon_{K}$ is very similar to the SM one. Therefore, it is
natural to expect some impact on the unitarity triangle fit, i.e. in
the SM CP--violating phase $\phi_1$. More precisely, in the SM the
constraint from $\varepsilon_{K}$ selects only positive values of
$\eta$ and hence constrains $\beta$ to be in the range 
$0 \leq \beta \leq \pi/2$. In
this model, the new contributions modify slightly this picture, but
they still fix a minimal value of $\beta$. This constraint is new with respect
to the analysis presented in \cite{EyalNir}.

In the $B$ sector, the relevant constraints come from , $\Delta M_{B_{d}}$, 
$\Delta M_{B_{s}}$ and $B\rightarrow Xl^{+}l^{-}$. For $\Delta M_{B_{j}}$ we
have \cite{BB} 
\begin{eqnarray}
&\Delta M_{B_{j}} = \Frac{G_{F}^{2}M_{W}^{2}
\eta_{B_{j}}B_{B_{j}}f_{B_{j}}^{2}m_{B_{j}}}{6\pi ^{2}} S_{0}\left(
x_{t}\right)\ \left|{\lambda _{t}^{bj}}^{2} \Delta_{bj}\right|\nonumber \\ &\Delta _{bj}\ =\ 1\ -\ 3.3\ \Frac{U_{bj}}{\lambda
_{t}^{bj}}\ +\ 165\ \left( \Frac{U_{bj}} {\lambda _{t}^{bj}}\right)^{2},
\label{deltam}
\end{eqnarray}
where the new parameters are defined in Ref.~\cite{BurasH}, and the
experimental values are $\Delta M_{B_{d}}=\left( 0.472\pm 0.017\right)
\times 10^{12}\ \mbox{s}^{-1}$ and $\Delta M_{B_{s}}>10.6\times
10^{12}\ \mbox{s}^{-1}$. From the upper bound on $B\rightarrow X_{s}l^{+}l^{-}$
\cite {cleo} we have\cite{BurasH,BHI}
\begin{equation}
\left| Y_{0}\left( x_{t}\right) \lambda _{t}^{bs}\ +\ C_{U2Z}\
U_{bs}\right| <0.15.
\label{UBS}
\end{equation}
Note that the SM prediction is much below the actual experimental bound; 
therefore, in order to constrain $U_{bs}$ it is enough to include the
leading SM contribution (the one with $Y_{0}(x_{t})$), the leading new
physics one, and their interference. Other subleading pieces \cite{BurasH}
have been neglected in Eq.~(\ref{UBS}). For
completeness, we recall that the bound $\left| U_{bd}\right|
<1.6\times 10^{-3}$ is obtained from $B\rightarrow X_{d}l^{+}l^{-}$,
neglecting the SM contribution. Nevertheless, this bound is not relevant 
once the constraint from $\Delta M_{B_{d}}$ is included.

To find the allowed region in the 9--dimensional parameter space of
the matrix $V$, we impose the 95\% C.L. experimental constraints and
we treat hadronic uncertainties as independent theoretical errors at
$1\sigma $. The important quantities to signal new physics in these
models are the FCNC couplings $U_{ds}$, $U_{bd}$ and $U_{bs}$. In a
first analysis we leave aside the $a_{J/\psi}$ constraint.

\begin{figure}
\begin{center}
\epsfxsize= \linewidth
\epsfbox{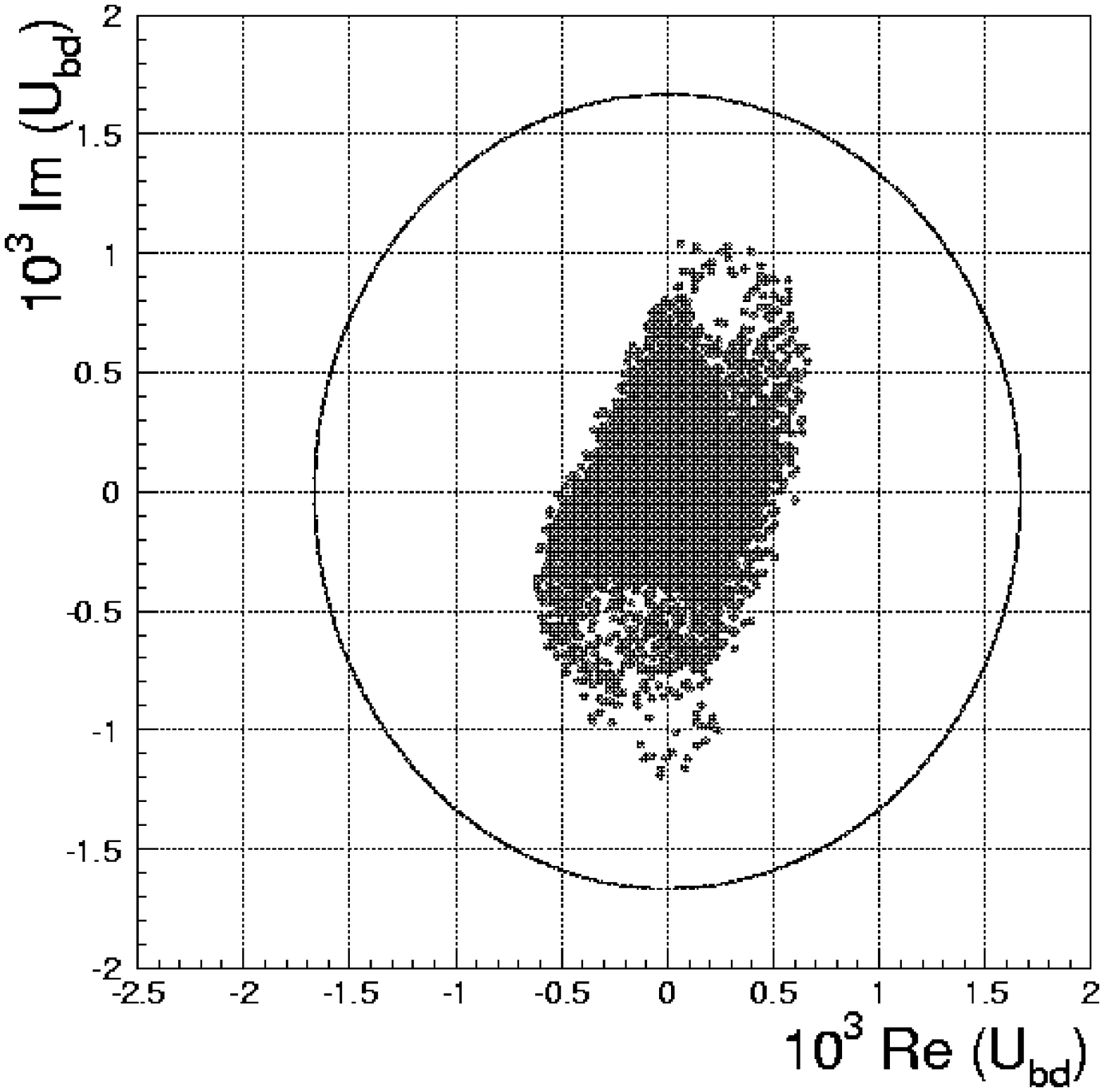}
\epsfxsize= \linewidth
\epsfbox{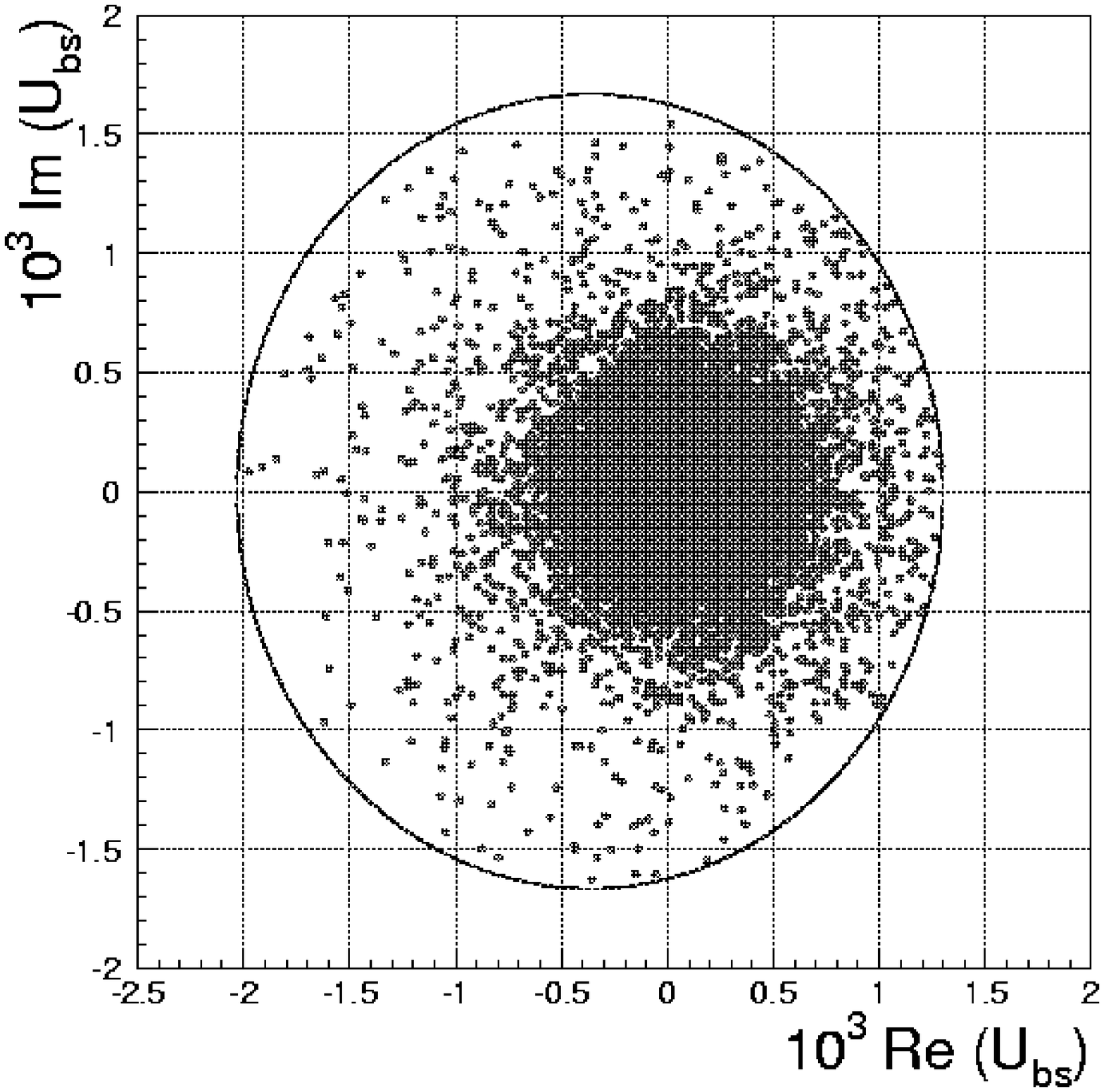}
\end{center}
\vskip 0.1cm
\caption{Scatter plot of the allowed $U_{bd}$ and $U_{bs}$ with all the 
constraints described in the text, but no $B^0$ CP asymmetry requirement.}
\label{fig:noasim}
\end{figure}

Taking $B_{6}^{(1/2)}=1.3 \pm 0.5$ (the case where the SM calculation
includes the experimental result of $\varepsilon^{\prime}/\varepsilon$), 
we get an approximate rectangular region in
the plane $U_{ds}$: $-3\times 10^{-6}\lesssim \mbox{Re}\left(
U_{ds}\right) \lesssim 4\times 10^{-6}$ and $-1.7\times 10^{-6}\lesssim
\mbox{Im}\left( U_{ds}\right) \lesssim 5.5\times 10^{-6}$. These bounds
turn to be a factor 2 better than the bounds usually quoted in the
literature, because of the inclusion of all the different correlations
by using a complete parametrization for $V$. For such small values
of $U_{ds}$, the $\varepsilon _{K}$ expression is similar to the SM one,
and hence a bound on $\gamma \simeq \phi _{1}$, the SM CP--violating phase, 
is also obtained. In order to fulfill the $\varepsilon_{K}$ constraint, we get
$0.6\lesssim \phi _{1}\lesssim 3$. Moreover, with the help of the unitarity
quadrangle \cite{BBBV}, including the general bound on $U_{bd}$, we get
also $-0.06\lesssim \beta \lesssim 0.6$, a bigger range than in the
SM model but in any case essentially positive \cite{EyalNir}. Notice
that for low $U_{bd}$, the correlation between $\beta $ and $\phi_{1}$
is similar to the usual one in the SM analysis of the unitarity
triangle. In Fig.~\ref{fig:noasim},
, we present a complete scatter plot
for $U_{bd}$ and $U_{bs}$ varying all the angles and phases in their 
allowed ranges and imposing all the constraints discussed above. 
As we can see in the $U_{bd}$ plot, 
we obtain $|U_{bd}| \leq 1.2 \times 10^{-3}$, which is controlled by
the $\Delta M_{B_{d}}$ upper bound \cite{BBBV,BEYALNir}. To set a 
reference scale, we include in the figure the circle corresponding 
to the $B\rightarrow X_{d}l^{+}l^{-}$ bound which, noticeably, is 
only a factor $\sqrt{2}$ above the final upper bound. In the
$U_{bs}$ plane, the lower bound on $\Delta M_{B_{s}}$ does not
fix an upper value for $|U_{bs}|$, and this is controlled by the curve 
from Eq.~(\ref{UBS}), i.e. $B\rightarrow X_{s}l^{+}l^{-}$ is the
relevant bound, which roughly fixes $|U_{bs}| \leq 2 \times 10^{-3}$.

If we use $B_{6}^{(1/2)}=1\pm 0.2$ to perform the analysis, no
relevant changes appear in Fig.~\ref{fig:noasim}
, that is, at this level the bounds on $U_{bs}$ and $U_{bd}$ are 
not modified. Of course, the rectangle in the $U_{ds}$ plane 
changes its imaginary region to 
$1.9\times 10^{-7}\lesssim \mbox{Im}\left( U_{ds}\right) \lesssim
6.2\times 10^{-6}$, indicating the need of new physics for
$\varepsilon^{\prime}/\varepsilon$.

\begin{figure}
\begin{center}
\epsfxsize= \linewidth
\epsfbox{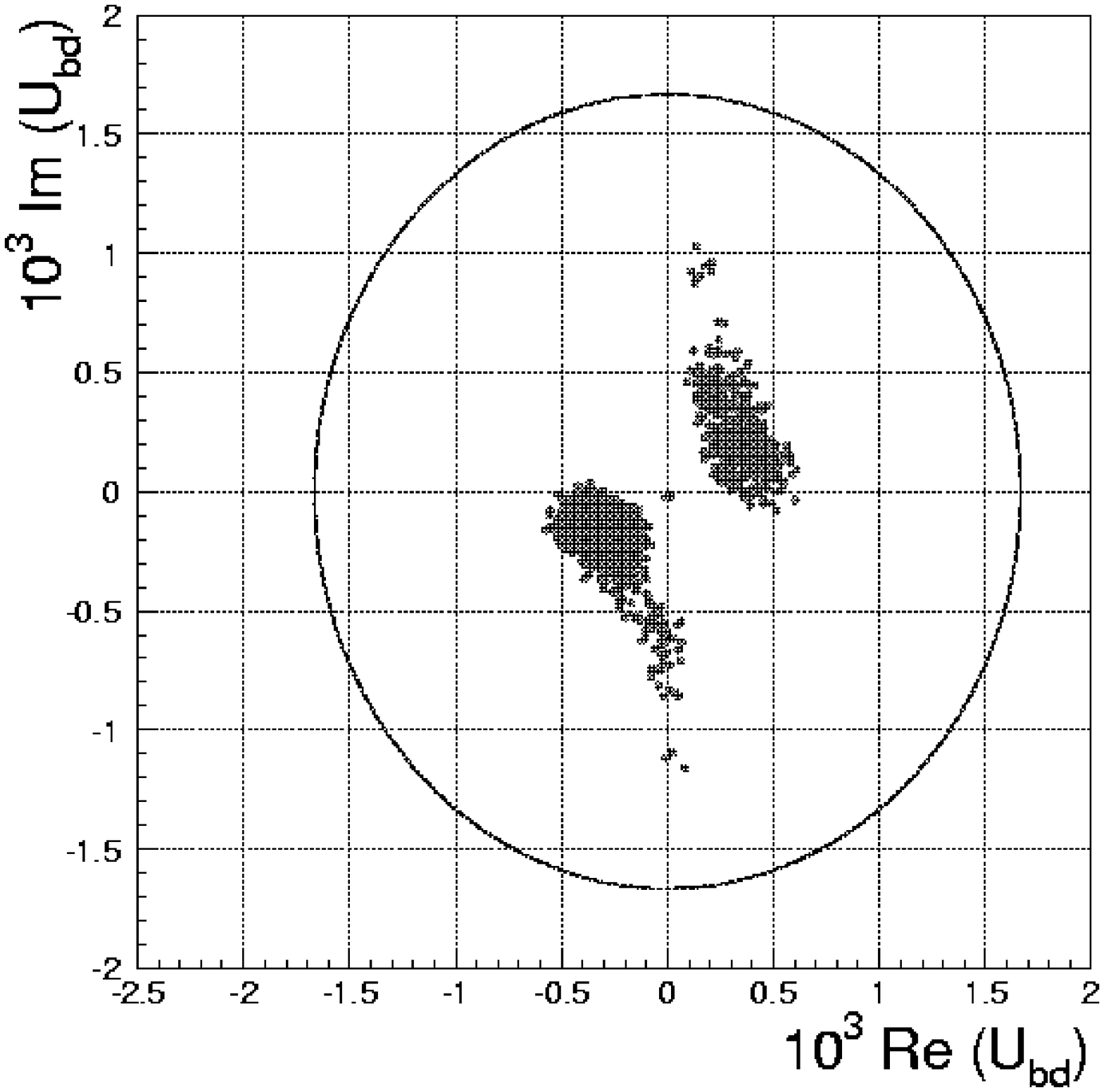}
\epsfxsize= \linewidth
\epsfbox{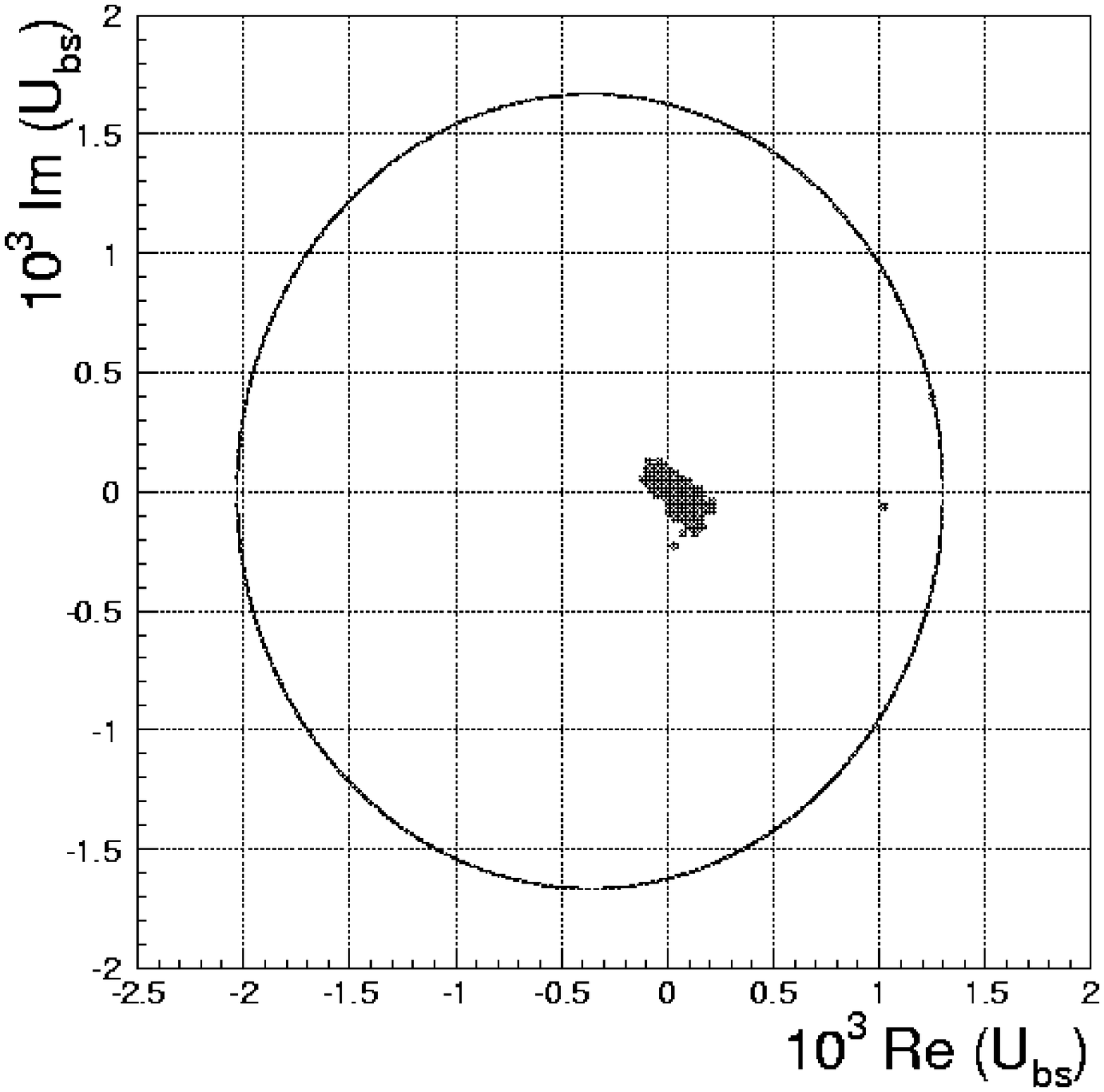}
\end{center}
\vskip 0.1cm
\caption{The same plot as before with the additional requirement on the 
$a_{J/\psi}$ CP asymmetry to reproduce the Babar value, 
$a_{J/\psi}= 0.34 \pm 0.20$.}
\label{fig:babar}
\end{figure}

In this model, the $B^0 \to J/\psi\ K_s$ CP asymmetry, $a_{J/\psi}$, is
given by
\begin{equation}
a_{J/\psi}=\sin \left( 2\beta -\arg \Delta _{bd}\right).   \label{ajkvl}
\end{equation}
In order to illustrate the effects of a low $a_{J/\psi}$ value, we
have incorporated to the previous analysis the Babar range:
$0.14\lesssim a_{J/\psi}\lesssim 0.54$. Figure~\ref{fig:babar}  
 shows the corresponding scatter plot for the $U_{bd}$ and $U_{bs}$
planes. It is important to emphasize that these plots are directly obtained
from Fig.~\ref{fig:noasim}, with the only additional 
requirement of the Babar asymmetry, that is, 
these points are only a subset of the allowed region in 
Fig.~\ref{fig:noasim}. Therefore, we can see here the very strong 
impact of this asymmetry both in 
the $U_{bd}$ and $U_{bs}$ couplings \cite{EyalNir}. 
From Fig.~\ref{fig:babar}
we see that, in the $U_{bd}$ plane, the great majority of the allowed points 
are in the range
$2\times 10^{-4}\lesssim \left| U_{bd}\right| \lesssim 1.2\times 10^{-3}$,
i.e. a large, non--vanishing $U_{bd}$ coupling is required to reproduce
the Babar asymmetry. In particular, this means that,
within this model, a low CP asymmetry  implies the presence of
new physics in the $B$ system, independently of the existence of 
non--vanishing contributions to the $K$ system ($U_{sd} \neq 0$). 
Concerning this, we must remember that, in principle, a low CP asymmetry
could also be due to a large new contribution in kaon physics with a
negligible contribution to the $B$ system \cite{smallajk} (see, in particular,
the last two references in \cite{smallajk} for an example of this).
However, as we have seen, in this model, the $\varepsilon_K$ constraint 
does not depart largely from the SM situation, and so, only a large $U_{bd}$
coupling can produce the required effect. Indeed, models with additional
vector--like quarks constitute the simplest extensions of the SM which
modify strongly the $B^0$ CP asymmetries through a new contribution in the 
$B$ system.
 
On the other hand, we see that, for these points, the coupling
$U_{bs}$ is always restricted to the range $|U_{bs}|\lesssim 2 \times
10^{-4}$; hence all the allowed points have simultaneously high
$\left|U_{bd}\right|$ and low $\left| U_{bs}\right|$.  Indeed, it is
easy to obtain, from Eq.~(\ref{ZFCNC}), the relation
$U_{bd}U_{bs}^{*}=- U_{sd}\left| V_{4b}\right|^{2}$.  The region in the
$U_{ds}$ plane does not change with the inclusion of the $a_{J/\psi}$
constraint, and then we still have, $|U_{sd}|\lesssim 6 \times
10^{-6}$ and $|V_{4b}|^2 \lesssim 0.009$. Taking into account that a
low $a_{J/\psi}$ requires $|U_{bd}| \geq 2 \times 10^{-4}$, this
clearly implies an absolute upper bound, $|U_{bs}|\lesssim 3 \times
10^{-4}$, that turns to be $\lesssim 10^{-4}$ when all the
correlations are included.  Therefore, for this set of points, we can
not expect a new--physics contribution in the $b \to s$ transition.
It is important to emphasize, once more, that these results are
independent of the existence of sizeable effects in the kaon system
and, in particular of the chosen value for $B_{6}^{(1/2)}$.

At this point, it is very interesting to examine the predicted branching
ratios of the decays $B\to X_{d,s} l \bar{l}$ for this set of points. 
From Fig.~\ref{fig:babar}, where 
we have included the circle corresponding to the experimental
bounds in these decays, it is clear that we can also expect a very
large contribution to $B\to X_{d} l \bar{l}$. In this case,
the branching ratios for the $X_d$ decays are strongly enhanced from the 
SM prediction, reaching values of $1.0 \times 10^{-6} \leq 
BR\left(B\rightarrow X_{d}l^{+}l^{-}\right) \leq 1.8 \times 10^{-5}$ 
and $6.0 \times 10^{-5} \leq BR\left(B\rightarrow X_{d}\nu \bar{\nu}\right) 
\leq 1.0 \times 10^{-4}$. While, on the other hand, the low values
of $U_{bs}$ imply that the $X_s$ decays remain roughly at the SM 
value. 

In Figure~\ref{fig:babar}, we also find a few points ($\simeq 0.1 \%$
of the points) which have simultaneously $|U_{bs}|\gsim 1 \times
10^{-3}$ and $|U_{bd}|\lesssim 3 \times 10^{-5}$.  This second class
of points is only possible in the vicinity of the SM and they
disappear if the value of the asymmetry is reduced to
$a_{J/\psi}\lesssim 0.52$. Still, it is important to emphasize that
these points also require the presence of new physics in $B$
decays. In fact, although there is no sizeable departure from the SM
expectations in $B\to X_{d} l \bar{l}$, the $X_s$ decays are now close
to the experimental upper range. Namely, we obtain, for the point to
the right of Fig.~\ref{fig:babar}, with $Re\left(U_{bs}\right) \simeq
1 \times 10^{-4}$, $BR\left(B\rightarrow X_{s}l^{+}l^{-}\right) \simeq
2.7 \times 10^{-5} $ to be compared with the experimental upper bounds
of $BR\left(B\rightarrow X_{s}l^{+}l^{-}\right)\leq 4.2 \times
10^{-5}$.  However, this possibility is marginal in the $1 \sigma$
Babar range, and we do not discuss it any further here.

If the analysis is made with the world average, the
$U_{bs}$ scatter plot is very similar to the one of Fig.~\ref{fig:noasim}. 
The $U_{bd}$ plot changes significantly from Fig.~\ref{fig:noasim}. 
The outer regions in the second and fourth
quadrants are reduced and the central region corresponding to the SM remains
filled; this situation represents an improved version of the
analysis presented in Ref.~\cite{EyalNir}.

We have to conclude that, in the context of models with vector--like
singlet quarks, a low value of $a_{J/\psi}\lesssim 0.5$ implies the 
presence of FCNC in the $b \to d$ transition and its absence in 
$b \to s$ transitions. This is completely independent of the 
presence or absence of sizeable new--physics contributions in the kaon system. 
More importantly, an additional and clean signature of this
scenario would be a rather high value for the branching ratios of the
tree--level dominated rare decays: $B\rightarrow X_{d}l^{+}l^{-} $ and
$B\rightarrow X_{d}\nu \overline{\nu }$, with enhancement factors
${\cal{O}}(20)$ over the SM expectations.
\\

The work of F.B. and O.V. was partially supported by the Spanish CICYT
AEN-99/0692 and from the EU network ``Eurodaphne'', contract number
ERBFMRX-CT98-0169. O.V.  acknowledges financial support from a Marie Curie
EC grant (HPMF-CT-2000-00457).

% now the references. delete or change fake bibitem. delete next three
%   lines and directly read in your .bbl file if you use bibtex.

\newpage

\begin{references}

\bibitem{quico} 
F.~J.~Botella and L.~Chau,
%``Anticipating The Higher Generations Of Quarks From Rephasing Invariance Of The Mixing Matrix,''
Phys.\ Lett.\  {\bf B168}, 97 (1986).
%%CITATION = PHLTA,B168,97;%%

\bibitem{branco}
G.~C.~Branco and L.~Lavoura,
%``Rephasing Invariant Parametrization Of The Quark Mixing Matrix,''
Phys.\ Lett.\  {\bf B208}, 123 (1988);
%%CITATION = PHLTA,B208,123;%%
\\
G.~C.~Branco and L.~Lavoura,
%``Wolfenstein Type Parametrization Of The Quark Mixing Matrix,''
Phys.\ Rev.\  {\bf D38}, 2295 (1988).
%%CITATION = PHRVA,D38,2295;%%

\bibitem{babar} 
B.~Aubert {\it et al.} [Babar Collaboration], Report No. SLAC-PUB-8540,
%``A study of time-dependent CP-violating asymmetries in  B0 --> J/psi K0(S) and B0 --> psi(2S) K0(S) decays,''
hep-ex/0008048;
%%CITATION = HEP-EX 0008048;%%2000)
\\
B.~Aubert {\it et al.}  [BaBar Collaboration],
%``Measurement of CP violating asymmetries in B0 decays to CP eigenstates,''
Phys.\ Rev.\ Lett.\ {\bf 86}, 2515 (2001)
[hep-ex/0102030].
%%CITATION = HEP-EX 0102030;%%


\bibitem{belle} 
H.~Aihara [Belle Collaboration], Belle note 353,
%``A measurement of CP violation in B0 meson decays with Belle,''
hep-ex/0010008;
%%CITATION = HEP-EX 0010008;%%
\\
A.~Abashian {\it et al.}  [BELLE Collaboration],
%``Measurement of the CP violation parameter sin 2phi(1) in B/d0 meson  decays,''
Phys.\ Rev.\ Lett.\ {\bf 86}, 2509 (2001)
[hep-ex/0102018].
%%CITATION = HEP-EX 0102018;%%


\bibitem{CDF}
T.~Affolder {\it et al.}, CDF Collaboration,
%``A measurement of sin(2beta) from B --> J/psi K0(S) with the CDF  detector,''
Phys.\ Rev.\  {\bf D61}, 072005 (2000), hep-ex/9909003.
%%CITATION = HEP-EX 9909003;%%

\bibitem{smallajk} 
J.~P.~Silva and L.~Wolfenstein, Report No. SLAC-PUB-8548, 
%``Implications of the possibility that sin(2 beta) is small,''
hep-ph/0008004;
%%CITATION = HEP-PH 0008004;%%
\\
G.~Eyal, Y.~Nir and G.~Perez,
%``Implications of a small CP asymmetry in B --> psi K(S),''
JHEP {\bf 0008}, 028 (2000),
hep-ph/0008009;
%%CITATION = HEP-PH 0008009;%%
\\
Z.~Xing, 
%``Possible implications of small CP violation in B/d0 vs anti-B/d0  --> J/psi K(S) decays,''
hep-ph/0008018;
%%CITATION = HEP-PH 0008018;%%
\\
A.~J.~Buras and R.~Buras, Report No. TUM-HEP-285-00,
%``A lower bound on sin(2beta) from minimal flavor violation,''
hep-ph/0008273;
%%CITATION = HEP-PH 0008273;%%
\\
A.~Masiero and O.~Vives, Report No. SISSA-69-2000-EP, to be published in 
Phys.\ Rev.\ Lett.,
%``Kaon vs. bottom: Where to look for a general MSSM?,''
hep-ph/0007320;
%%CITATION = HEP-PH 0007320;%%
\\
A.~Masiero, M.~Piai and O.~Vives, Report No. FTUV-12-08,
%``Supersymmetric origin of a low $a_{J/\psi}$ CP asymmetry,''
hep-ph/0012096.
%%CITATION = HEP-PH 0012096;%%

\bibitem{vlq}
F.~del Aguila and J.~Cortes,
%``A New Model Of Weak CP Violation,''
Phys.\ Lett.\  {\bf B156}, 243 (1985);
%%CITATION = PHLTA,B156,243;%%
\\
G.~C.~Branco and L.~Lavoura,
%``On The Addition Of Vector Like Quarks To The Standard Model,''
Nucl.\ Phys.\  {\bf B278}, 738 (1986);
%%CITATION = NUPHA,B278,738;%%
\\
F.~del Aguila, M.~K.~Chase and J.~Cortes,
%``Vector Like Fermion Contributions To Epsilon-Prime,''
Nucl.\ Phys.\  {\bf B271}, 61 (1986);
%%CITATION = NUPHA,B271,61;%%
\\
Y.~Nir and D.~J.~Silverman,
%``Z Mediated Flavor Changing Neutral Currents And Their Implications For CP Asymmetries In B0 Decays,''
Phys.\ Rev.\  {\bf D42}, 1477 (1990);
%%CITATION = PHRVA,D42,1477;%%
\\
D.~Silverman,
%``Z mediated B - anti-B mixing and B meson CP violating asymmetries in the light of new FCNC bounds,''
Phys.\ Rev.\  {\bf D45}, 1800 (1992);
%%CITATION = PHRVA,D45,1800;%%
\\
G.~C.~Branco, T.~Morozumi, P.~A.~Parada and M.~N.~Rebelo,
%``CP asymmetries in B0 decays in the presence of flavor changing neutral currents,''
Phys.\ Rev.\  {\bf D48}, 1167 (1993);
%%CITATION = PHRVA,D48,1167;%%
\\
W.~Choong and D.~Silverman,
%``New phases in CP violating B decay asymmetries from mixing to singlet down quarks,''
Phys.\ Rev.\  {\bf D49}, 2322 (1994);
%%CITATION = PHRVA,D49,2322;%%
\\
V.~Barger, M.~S.~Berger and R.~J.~Phillips,
%``Quark singlets: Implications and constraints,''
Phys.\ Rev.\  {\bf D52}, 1663 (1995),
 hep-ph/9503204;
%%CITATION = HEP-PH 9503204;%%
\\
D.~Silverman,
%``B factory constraints on isosinglet down quark mixing, and predictions for other CP violating experiments,''
Int.\ J.\ Mod.\ Phys.\  {\bf A11}, 2253 (1996),
 hep-ph/9504387;
%%CITATION = HEP-PH 9504387;%%
\\
M.~Gronau and D.~London,
%``New physics in CP asymmetries and rare B decays,''
Phys.\ Rev.\  {\bf D55}, 2845 (1997),
 hep-ph/9608430;
%%CITATION = HEP-PH 9608430;%%
\\
F.~del Aguila, J.~A.~Aguilar-Saavedra and G.~C.~Branco,
%``CP violation from new quarks in the chiral limit,''
Nucl.\ Phys.\  {\bf B510}, 39 (1998),
 hep-ph/9703410.
%%CITATION = HEP-PH 9703410;%%

\bibitem{ChauKeung} 
L.~Chau and W.~Keung,
%``Comments On The Parametrization Of The Kobayashi-Maskawa Matrix,''
Phys.\ Rev.\ Lett.\  {\bf 53}, 1802 (1984).
%%CITATION = PRLTA,53,1802;%%

\bibitem{PDG}
D.~E.~Groom {\it et al.},
%``Review of particle physics,''
Eur.\ Phys.\ J.\  {\bf C15}, 1 (2000).
%%CITATION = EPHJA,C15,1;%%

\bibitem{LuisJoao}  
L.~Lavoura and J.~P.~Silva,
%``Bounds on the mixing of the down type quarks with vector - like singlet quarks,''
Phys.\ Rev.\  {\bf D47}, 1117 (1993).
%%CITATION = PHRVA,D47,1117;%%

\bibitem{Paco}
F.~del Aguila, J.~A.~Aguilar-Saavedra and R.~Miquel,
%``Constraints on top couplings in models with exotic quarks,''
Phys.\ Rev.\ Lett.\  {\bf 82}, 1628 (1999), hep-ph/9808400.
%%CITATION = HEP-PH 9808400;%%

\bibitem{BurasSilvestrini} 
A.~J.~Buras and L.~Silvestrini,
%``Upper bounds on K --> pi nu anti-nu and K(L) --> pi0 e+ e- from  epsilon'/epsilon and K(L) --> mu+ mu-,''
Nucl.\ Phys.\  {\bf B546}, 299 (1999),
 hep-ph/9811471.
%%CITATION = HEP-PH 9811471;%%
%\href{\wwwspires?eprint=HEP-PH/9811471}{SPIRES}

\bibitem{PichDumm} 
D.~Gomez Dumm and A.~Pich,
%``Long-distance contributions to the K(L) --> mu+ mu- decay width,''
Phys.\ Rev.\ Lett.\  {\bf 80}, 4633 (1998),
 hep-ph/9801298;
%%CITATION = HEP-PH 9801298;%%
\\
G.~D'Ambrosio, G.~Isidori and J.~Portoles,
%``Can we extract short-distance information from B(K(L) --> mu+ mu-)?,''
Phys.\ Lett.\  {\bf B423}, 385 (1998),
 hep-ph/9708326.
%%CITATION = HEP-PH 9708326;%%


\bibitem{I-L}
T.~Inami and C.~S.~Lim,
%``Effects Of Superheavy Quarks And Leptons In Low-Energy Weak Processes K(L) $\to$ Mu Anti-Mu, K+ $\to$ Pi+ Neutrino Anti-Neutrino And K0 <--->Anti-K0,''
Prog.\ Theor.\ Phys.\  {\bf 65}, 297 (1981).
%%CITATION = PTPKA,65,297;%%

\bibitem{BurasH}  
A.~J.~Buras and R.~Fleischer, Report No. TUM-HEP-275-97,
%``Quark mixing, CP violation and rare decays after the top quark  discovery,''
hep-ph/9704376.
%%CITATION = HEP-PH 9704376;%%

\bibitem{BurasMart}
M.~Ciuchini, E.~Franco, L.~Giusti, V.~Lubicz and G.~Martinelli,
Report No. ROME-99-1268,
%``epsilon'/epsilon from lattice QCD,''
hep-ph/9910237;
%%CITATION = HEP-PH 9910237;%%
\\
M.~Ciuchini and G.~Martinelli, Report No. TUM-HEP-376-00, 
%``Theoretical status of epsilon'/epsilon,''
hep-ph/0006056;
%%CITATION = HEP-PH 0006056;%%
\\
S.~Bosch, A.~J.~Buras, M.~Gorbahn, S.~Jager, M.~Jamin, M.~E.~Lautenbacher 
and L.~Silvestrini,
%``Standard model confronting new results for epsilon'/epsilon,''
Nucl.\ Phys.\  {\bf B565}, 3 (2000),
 hep-ph/9904408.
%%CITATION = HEP-PH 9904408;%%

\bibitem{Bertolini} 
S.~Bertolini, J.~O.~Eeg, M.~Fabbrichesi and E.~I.~Lashin,
%``epsilon'/epsilon at O(p**4) in the chiral expansion,''
Nucl.\ Phys.\  {\bf B514}, 93 (1998),
 hep-ph/9706260;
%%CITATION = HEP-PH 9706260;%%
\\
S.~Bertolini, M.~Fabbrichesi and J.~O.~Eeg,
%``Theory of the CP-violating parameter epsilon'/epsilon,''
Rev.\ Mod.\ Phys.\  {\bf 72}, 65 (2000),
 hep-ph/9802405.
%%CITATION = HEP-PH 9802405;%%

\bibitem{PichPallante} 
E.~Pallante and A.~Pich,
%``Strong enhancement of epsilon'/epsilon through final state  interactions,''
Phys.\ Rev.\ Lett.\  {\bf 84}, 2568 (2000),
 hep-ph/9911233.
%%CITATION = HEP-PH 9911233;%%

\bibitem{Nirnew} 
Y.~Nir, Report No. IASSNS-HEP-99-96,
%``CP violation in and beyond the standard model,''
%in {\it NONE}
hep-ph/9911321.
%%CITATION = HEP-PH 9911321;%%

\bibitem{BB} 
G.~Barenboim and F.~J.~Botella,
%``Delta(F) = 2 effective Lagrangian in theories with vector-like  fermions,''
Phys.\ Lett.\  {\bf B433}, 385 (1998),
 hep-ph/9708209.
%%CITATION = HEP-PH 9708209;%%

\bibitem{EyalNir}  
G.~Eyal and Y.~Nir,
%``Constraining extensions of the quark sector with the CP asymmetry in  B --> psi K(S),''
JHEP {\bf 9909}, 013 (1999),
 hep-ph/9908296.
%%CITATION = HEP-PH 9908296;%%

\bibitem{cleo} 
S.~Glenn {\it et al.}  [CLEO Collaboration],
%``Search for inclusive b --> s l+ l-,''
Phys.\ Rev.\ Lett.\  {\bf 80}, 2289 (1998),
 hep-ex/9710003.
%%CITATION = HEP-EX 9710003;%%


\bibitem{BHI} 
G.~Buchalla, G.~Hiller and G.~Isidori, Report No. SLAC-PUB-8430, 
%``Phenomenology of non-standard Z couplings in exclusive semileptonic  b --> s transitions,''
hep-ph/0006136.
%%CITATION = HEP-PH 0006136;%%

\bibitem{BBBV} 
G.~Barenboim, F.~J.~Botella, G.~C.~Branco and O.~Vives,
%``How sensitive to FCNC can B0 CP asymmetries be?,''
Phys.\ Lett.\  {\bf B422}, 277 (1998),
hep-ph/9709369.
%%CITATION = HEP-PH 9709369;%%


\bibitem{BEYALNir}
G.~Barenboim, G.~Eyal and Y.~Nir,
%``Constraining new physics with the CDF measurement of CP violation in  B --> psi K(S),''
Phys.\ Rev.\ Lett.\  {\bf 83}, 4486 (1999),
hep-ph/9905397.
%%CITATION = HEP-PH 9905397;%%

\end{references}
\end{document}